\pdfoutput=1
\documentclass[11pt,a4paper,twoside,openright,closeany,textany]{book}
\usepackage{makeidx}
\usepackage{graphicx}
\usepackage{paralist}
\usepackage{caption} 
\usepackage{amsmath}
\usepackage{amssymb}
\usepackage{mathtools}
\usepackage{fancyhdr}
\usepackage{multicol}
\usepackage{multirow}

\usepackage{mdframed}
\usepackage[usenames,dvipsnames,svgnames,table]{xcolor}
\usepackage{array}
\usepackage[%
  colorlinks=true,%
  linkcolor=black,%
  urlcolor=black,%
  citecolor=black%
]{hyperref}
\usepackage{nameref} 
\usepackage{ifthen} 
\usepackage[font={small,sl}]{caption}
\usepackage[authoryear]{natbib}
\renewcommand\bibname{References}
\newcommand{\mychapbib}{
  \addcontentsline{toc}{section}{\bibname}
  \bibliographystyle{natbib}
  \bibliography{strucbioinf}
}
\usepackage[colorinlistoftodos]{todonotes}
\usepackage{letltxmacro}

\usepackage{tocstyle} 
\usetocstyle{standard}
\settocfeature{raggedhook}{\raggedright}
\newcommand{\bibfont}{\small\raggedright}
\def\cite{\citep}
\newcommand{\citeeg}[1]{\cite[e.g.,][]{#1}}

\LetLtxMacro{\oldTodo}{\todo}
\renewcommand{\todo}[2][]{\oldTodo[#1]{TODO: #2}}

\usepackage[normalem]{ulem}

\newcommand\inwish[1]{\oldTodo[inline,color=SkyBlue]{WISH: #1}}

\newboolean{onechapter}

\newcommand{\AF}[1][~]{K.\@#1Anton#1Feenstra}
\newcommand{\SA}[1][~]{Sanne#1Abeln}
\newcommand{\JH}[1][~]{Jaap#1Heringa}

\newcommand{\PB}[1][~]{Punto#1Bawono}
\newcommand{\BS}[1][~]{Bas#1Stringer}

\newcommand{\JvG}[1][~]{Juami#1H.\@#1M.\@#1van#1Gils}

\newcommand{\MD}[1][~]{Maurits#1Dijkstra}

\newcommand{\RB}[1][~]{\mbox{Robbin}#1\mbox{Bouwmeester}}

\newcommand{\JG}[1][~]{\mbox{Jose}#1\mbox{Gavald\'a-Garc\'ia}}

\newcommand{\MO}[1][~]{\mbox{Mascha}#1\mbox{Okounev}}
\newcommand{\DG}[1][~]{\mbox{Dea}#1\mbox{Gogishvili}}

\newcommand{\KW}[1][~]{\mbox{Katharina}#1\mbox{Waury}}

\newcommand{\TL}[1][~]{\mbox{Ting}#1\mbox{Liu}}
\newcommand{\IH}[1][~]{\mbox{Isabel}#1\mbox{Houtkamp}}
\newcommand{\CG}[1][~]{\mbox{Carlos}#1\mbox{Fernandez}#1\mbox{Garcia}}

\newcommand{\orcid}[1]{\href{https://orcid.org/#1}{\raisebox{-0.7ex}{\protect\includegraphics[height=3ex]{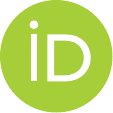}}}}
\definecolor{idgreen}{RGB}{166 206 57}
\newcommand{\mailid}[1]{\href{mailto:#1}{\raisebox{-0.3ex}{\color{idgreen}\textsf{\textbf{\Large \protect@}}}}}

\newcommand{\AFid}{\orcid{0000-0001-6755-9667}}
\newcommand{\SAid}{\orcid{0000-0002-2779-7174}}

\newcommand{\JGid}{\orcid{0000-0001-6431-3442}}
\newcommand{\JvGid}{\orcid{0000-0003-3706-7818}}

\newcommand{\DGid}{\orcid{0000-0001-8809-0861}}

\newcommand{\RBid}{\orcid{0000-0001-6807-7029}}

\newcommand{\KWid}{\orcid{0000-0002-8570-7640}}
\newcommand{\JHid}{\orcid{0000-0001-8641-4930}}

\newcommand{\BSid}{\orcid{0000-0001-7792-9385}}

\newcommand{\MDid}{\orcid{0000-0002-7971-6209}}

\newcommand{\TLid}{\orcid{0000-0002-8208-3430}}
\newcommand{\IHid}{\orcid{0000-0002-4222-7292}}
\newcommand{\CGid}{\orcid{0009-0000-1744-7325}}

\newcommand{\MOid}{\mailid{mascha.okounev@gmail.com}}
\newcommand{\PBid}{\mailid{puntobawono@gmail.com}}

\newcommand{\ACtxt}{Wrote the text}
\newcommand{\ACfig}{Created figures}
\newcommand{\ACref}{Review of current literature}
\newcommand{\ACeds}{Editorial responsibility}
\newcommand{\ACproof}{Critical proofreading}
\newcommand{\ACfb}{Non-expert feedback}

\newcommand{\Angs}[1][~]{\text{\normalfont\AA}}

\renewcommand{\and}{\quad}

\newcommand{\figlab}[1]{\textsf{\textbf{\Large #1}}}
\newcommand{\pdbref}[1]{\href{http://www.rcsb.org/pdb/explore.do?structureId=#1}{PDB:#1}}
\newcommand{\arxiv}[2][UNDEFINED]{\href{https://arxiv.org/abs/#2}{\ifthenelse{\equal{#1}{UNDEFINED}}{arxiv.org/abs/#2}{#1}}}

\newcommand{\figref}[2][]{\hyperref[fig:#2]{Figure\@~\ref*{fig:#2}#1}}
\newcommand{\tabref}[1]{\hyperref[tab:#1]{Table \ref*{tab:#1}}}
\renewcommand{\eqref}[2][]{\hyperref[eq:#2]{Equation#1\@~\ref*{eq:#2}}}
\newcommand{\panelref}[2][]{%
    \ifthenelse{\boolean{onechapter}}{%
        \hyperref[panel:#2]{Panel\@~``\nameref{panel:#2}#1''}%
    }{%
        \hyperref[panel:#2]{Panel\@~\ref*{panel:#2}#1}%
    }%
}
\newcommand{\secref}[2][n]{%
    \hyperref[sec:#2]{%
        \ifthenelse{\equal{#1}{n} }{Section\@~\ref*{sec:#2}}{}% just number
        \ifthenelse{\equal{#1}{nn}}{Section\@~\ref*{sec:#2} ``\nameref{sec:#2}''}{}% nm & nr
        \ifthenelse{\equal{#1}{N} }{``\nameref{sec:#2}''}{}% just quoted name
        \ifthenelse{\equal{#1}{NN} }{\nameref{sec:#2}}{}% just name
    }%
}
\newcommand{\chref}[2][n]{%
    \ifthenelse{\boolean{onechapter}}{%
        \ifthenelse{\equal{#2}{ChPref}     }{\arxiv[Chapter \ref*{ch:#2} ``\nameref*{ch:#2}'']{1801.09442}}{}%
        \ifthenelse{\equal{#2}{ChIntroPS}  }{\arxiv[Chapter \ref*{ch:#2} ``\nameref*{ch:#2}'']{2307.02169}}{}%
        \ifthenelse{\equal{#2}{ChDetVal}   }{\arxiv[Chapter \ref*{ch:#2} ``\nameref*{ch:#2}'']{2108.02706}}{}%
        \ifthenelse{\equal{#2}{ChStrucAli} }{\arxiv[Chapter \ref*{ch:#2} ``\nameref*{ch:#2}'']{2307.02170}}{}%
        \ifthenelse{\equal{#2}{ChDBClass}  }{\arxiv[Chapter \ref*{ch:#2} ``\nameref*{ch:#2}'']{2307.02171}}{}%
        \ifthenelse{\equal{#2}{ChFunc}     }{\arxiv[Chapter \ref*{ch:#2} ``\nameref*{ch:#2}'']{1801.09442}}{}%
        \ifthenelse{\equal{#2}{ChIntroPred}}{\arxiv[Chapter \ref*{ch:#2} ``\nameref*{ch:#2}'']{1712.00407}}{}%
        \ifthenelse{\equal{#2}{ChHomMod}   }{\arxiv[Chapter \ref*{ch:#2} ``\nameref*{ch:#2}'']{1712.00425}}{}%
        \ifthenelse{\equal{#2}{ChSSPred}   }{\arxiv[Chapter \ref*{ch:#2} ``\nameref*{ch:#2}'']{1801.09442}}{}
        \ifthenelse{\equal{#2}{ChFuncPred} }{\arxiv[Chapter \ref*{ch:#2} ``\nameref*{ch:#2}'']{2307.02173}}{}%
        \ifthenelse{\equal{#2}{ChIntroDyn} }{\arxiv[Chapter \ref*{ch:#2} ``\nameref*{ch:#2}'']{2307.02174}}{}%
        \ifthenelse{\equal{#2}{ChThermo}   }{\arxiv[Chapter \ref*{ch:#2} ``\nameref*{ch:#2}'']{2307.02175}}{}%
        \ifthenelse{\equal{#2}{ChMD}       }{\arxiv[Chapter \ref*{ch:#2} ``\nameref*{ch:#2}'']{2307.02176}}{}%
        \ifthenelse{\equal{#2}{ChMC}       }{\arxiv[Chapter \ref*{ch:#2} ``\nameref*{ch:#2}'']{2307.02177}}{}%
    }{
    \hyperref[ch:#2]{%
        \ifthenelse{\equal{#1}{n} }{Chapter \ref*{ch:#2}}{}% just number
        \ifthenelse{\equal{#1}{nn}}{Chapter \ref*{ch:#2} ``\nameref{ch:#2}''}{}% name & number
        \ifthenelse{\equal{#1}{N} }{``\nameref{ch:#2}''}{}% just name
      }%
  }%
}
\newcommand{\chrefname}[1]{\hyperref[ch:#1]{Chapter \ref*{ch:#1} ``\nameref{ch:#1}''}}
\newcommand{\partref}[1]{\hyperref[#1]{Part \ref*{#1}}}
\newcommand{\appref}[1]{\hyperref[app:#1]{Appendix \ref*{app:#1}}}
\newcommand{\footurl}[1]{\protect\footnote{\tt\href{https://#1}{#1}}}

\newcommand{\figsource}[1]{\protect\footnote{Figure source location: \url{#1}}}

\newenvironment{penum}[1][\itshape i)\upshape]
{\begin{inparaenum}[#1]} {\end{inparaenum}}

\renewcommand{\arraystretch}{1.3}

\makeatletter \@beginparpenalty=5000 \makeatother

\newenvironment{bgreading}[1][]{
  \begin{mdframed}[%
      outerlinewidth=0,%
      linecolor=CornflowerBlue!30,%
      backgroundcolor=CornflowerBlue!30,%
      innerleftmargin=14,%
      innerrightmargin=14,%
    ]
	\ifthenelse{\equal{#1}{}}{}{
        \stepcounter{panel}
    	\subsection*{#1} 
    }
}{%
  \end{mdframed}
}

\usepackage{listings}
\definecolor{backcolour}{rgb}{0.95,0.95,0.92}
\definecolor{codegreen}{rgb}{0,0.6,0}
\definecolor{codegray}{rgb}{0.5,0.5,0.5}
\definecolor{codered}{rgb}{0.8,0,0.0}
\definecolor{codeblue}{rgb}{0.0,0,0.8}
\lstdefinestyle{codeStyle}{
    backgroundcolor=\color{backcolour},   
    commentstyle=\color{codegreen},
    keywordstyle=\color{codeblue},
    numberstyle=\tiny\color{codegray},
    stringstyle=\color{codegray},
    numbers=left,                    
    tabsize=2
} 
\newcommand{\code}[1]{\texttt{#1}}

\makeindex

\begin{document}

\setboolean{onechapter}{true}

\pagestyle{fancy}
\lhead[\small\thepage]{\small\sf\nouppercase\rightmark}
\rhead[\small\sf\nouppercase\leftmark]{\small\thepage}
\newcommand{\innerfoot}{\footnotesize{\sf{\copyright} Feenstra \& Abeln}, 2014-2023}
\newcommand{\outerfoot}{\footnotesize \sf Intro Prot Struc Bioinf}
\lfoot[\outerfoot]{\innerfoot}
\cfoot{}
\rfoot[\innerfoot]{\outerfoot}
\renewcommand{\footrulewidth}{\headrulewidth}

\mainmatter
\setcounter{chapter}{8}
\chapterauthor{\MD~\MDid \and \KW~\KWid \and \DG~\DGid \and \PB~\PBid \and \IH~\IHid \and \JG~\JGid \and \MO~\MOid \and \RB~\RBid \and \BS~\BSid \and \JH~\JHid \and \SA*~\SAid \and \AF*~\AFid \and \JvG*~\JvGid}
\chapterfootnote{* editorial responsability}
\chapter{Structural Property Prediction}
\label{ch:ChSSPred}

\ifthenelse{\boolean{onechapter}}{\tableofcontents\newpage}{}

\section{Introduction}

The previous two chapters (\chref[n]{ChIntroPred} and \chref[n]{ChHomMod}) have shown us that predicting the three dimensional structure of a protein molecule from its amino acid sequence has been largely solved in recent years, although some challenges remain \cite{Liu2021}. Some structural properties, however, may be much easier to predict from sequence. Like tertiary structure, structural properties such as secondary structure, surface accessibility, flexibility and disorder, may be more strongly conserved than the primary sequence. Serving as building blocks for the native protein fold, these structural properties also contain important structural and functional information not apparent from the amino acid sequence directly. 

There are a few major reasons why structural property prediction is still a complicated task: \begin{penum} 
\item the large fraction of the structural data used to train various machine learning models is coming from static X-ray crystallography studies, however, globular proteins are dynamic and static information does not capture its characteristics completely; and 
\item proteins are part of a complex living system and do not exist in isolation: they may undergo various post-translational modifications, interactions or environmental alterations, leading to conformational changes not taken into account when considering it sequence or structure without context.
\end{penum}%
As such, knowledge of structural properties of a protein can contribute to tasks like fold recognition, but also be useful for multiple sequence alignment to find distant homologs, analysis of protein stability, and more generally for function prediction. In the next chapter (\chref[n]{ChFuncPred}) we will return to function prediction.

Here, we will first give an introduction into the application of machine learning for structural property prediction, and explain the concepts of cross-validation and benchmarking. Subsequently, we will discuss major concepts that play a key role in the characterization and prediction of structural properties: 
\begin{penum}
\item patterns of hydrophobicity along the amino acid sequence that relate to the three dimensional fold, 
\item the patterns in hydrogen bonding observed in $\alpha$-helices and $\beta$-sheets, 
\item intrinsic preferences of different amino acids to be in certain types of structural environments, and
\item evolutionary information as can be captured in sequence profiles.
\end{penum}%
Next, we will review various methods that incorporate knowledge of these concepts to predict those structural properties, such as secondary structure, surface accessibility, disorder and flexibility, and aggregation.
For an overview of more practical points to consider when developing such methods, please refer to ``Ten quick tips for sequence-based prediction of protein properties using machine learning'' by \citet{Hou2022}, and references therein.

\begin{figure}[b]
\centerline{
  \includegraphics[width=0.9\linewidth]{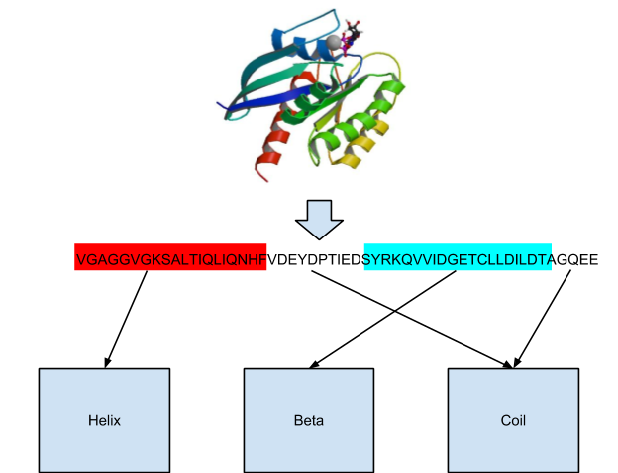}
}
\caption{Secondary structure prediction as a classification problem. Each of the residues in a protein sequence will be classified as being either of the prediction classes -- helix, strand or coil. Structure information (top part) is used as a reference for (supervised) learning, and as a gold standard for testing the accuracy of the predictions.}
\label{fig:ChSSPred-classification}
\end{figure}

\section{Structural property prediction as a machine learning problem}

Structural property prediction can be approached as a supervised machine learning problem; the aim here is to find patterns in the data that can explain the associated outcome. For this aim labeled data to learn on is required that already contains the true outputs, the \emph{labels}. Supervised learning can be divided further: Prediction of a specific \emph{class}, for instance the secondary structural component, as shown in \figref{ChSSPred-classification} is a \emph{classification} task. The prediction of a continuous \emph{value} such as disorder or solvent accessibility is a \emph{regression} task. The output of a supervised machine learning method is a predictor or model which allows us to predict the classes or values of an outcome variable. For a more in-depth explanation, please refer to \panelref{ChSSPred:ML}. 
Note that the predictors mentioned in this chapter, predict \emph{propensities} of amino acids to be part of a certain structural component such as: $\alpha$-helix, $\beta$-sheet or coil, protein-protein interaction (PPI) interface site, epitope or a hydrophobic patch. They generally do not give yes/no answers as output.

Most algorithms make structural property predictions per amino acid residue, such as NetSurfP-2.0 ~\citep{Klausen2019}. 
However, from the previous section you may already realise that we cannot predict the structural property of a residue in isolation: we need information on the surrounding residues, and potentially evolutionary conservation profiles around these positions in order to obtain accurate predictions. These, and other known properties at protein or residue level that carry useful information, are referred to as \emph{features} -- they will be the inputs to the model.

\begin{bgreading}[Key concepts and typical tasks in machine learning]
\label{panel:ChSSPred:ML}

We often use machine learning algorithms to try to increase our understanding of complex biological problems. In this box, we introduce some basic terminology in machine learning for those that have little experience in this field.

When we pose a biological question, we are often interested in what characteristics are specific to a certain group of samples, or in other words, what separates one group from another. In this case, we usually have a certain number of samples for each of the groups that we want to compare. Usually, the more samples we have per subgroup the better, as it may enable us to better separate biological signal from technological and biological noise.

Machine learning algorithms can broadly be differentiated based on whether or not an algorithm requires a ground truth to find patterns in the data:
\begin{compactitem}
\item[\bf Supervised learning:] When we train a model between groups for which the ground truth values (class label or continuous value) are known in advance, we are doing supervised learning. When we apply a supervised learning method, the data is split up into a training and a test set. The training set is used to train the model, whereas the test set is used to assess the performance of the model on data that it has not encountered before. This is necessary to avoid overfitting of the model. To prevent biased predictions, for most machine learning methods, the split in train and test should be balanced (similar percentage of labels or distribution of continuous values in train and test sets). 
\item[\bf Unsupervised learning:] When our aim is to identify interesting patterns in the data without prior knowledge about subgroups or correlations, we use unsupervised learning. Examples are principle component analysis (PCA), which calculates a weighted combination of variables that explains the largest variation in the dataset, and hierarchical clustering, which describes the similarity between samples using features of the dataset.
\end{compactitem}

\noindent
Supervised machine learning tasks commonly belong to one of the two following approaches depending on the type of outcome to predict:
\begin{compactitem}
\item[\bf classification:] when we try to classify samples into known classes, we want to predict the class labels from known variables. The variables in the dataset are called \emph{features}, and the process of identifying the most relevant features to be used for a particular prediction task is called \emph{feature selection}. Features can be either \emph{continuous} (e.g.\@ percentage of aromatic amino acids) or \emph{categorical} (e.g.\@ known binding to RNA). Although the input features can be continuous or categorical, the output is always categorical.
\item[\bf regression:] Alternatively, we may be interested in the relationship between some continuous features. For example, we might ask whether protein length is related to protein aggregation likelihood. To answer such questions, we use \emph{regression} models. The most simplest is \emph{linear regression}, which approximates a relationship between two variables with a linear model. If the slope of this fit is significantly different from zero, we can say that there is an association between the two variables. The prediction model then uses combinations of such linear fits to predict the value of the output variable from the input features. 
\item[\bf multi-task:] Moreover, we could be interested in classification and/or regression of multiple related output variables given one set of input features. This is called multi-task learning. With such methods we could for example predict secondary structural elements (classification), surface accessibility (classification or regression) residues and disorder (regression) simultaneously. 
\end{compactitem}

 \centerline{\includegraphics[width=0.9\linewidth]{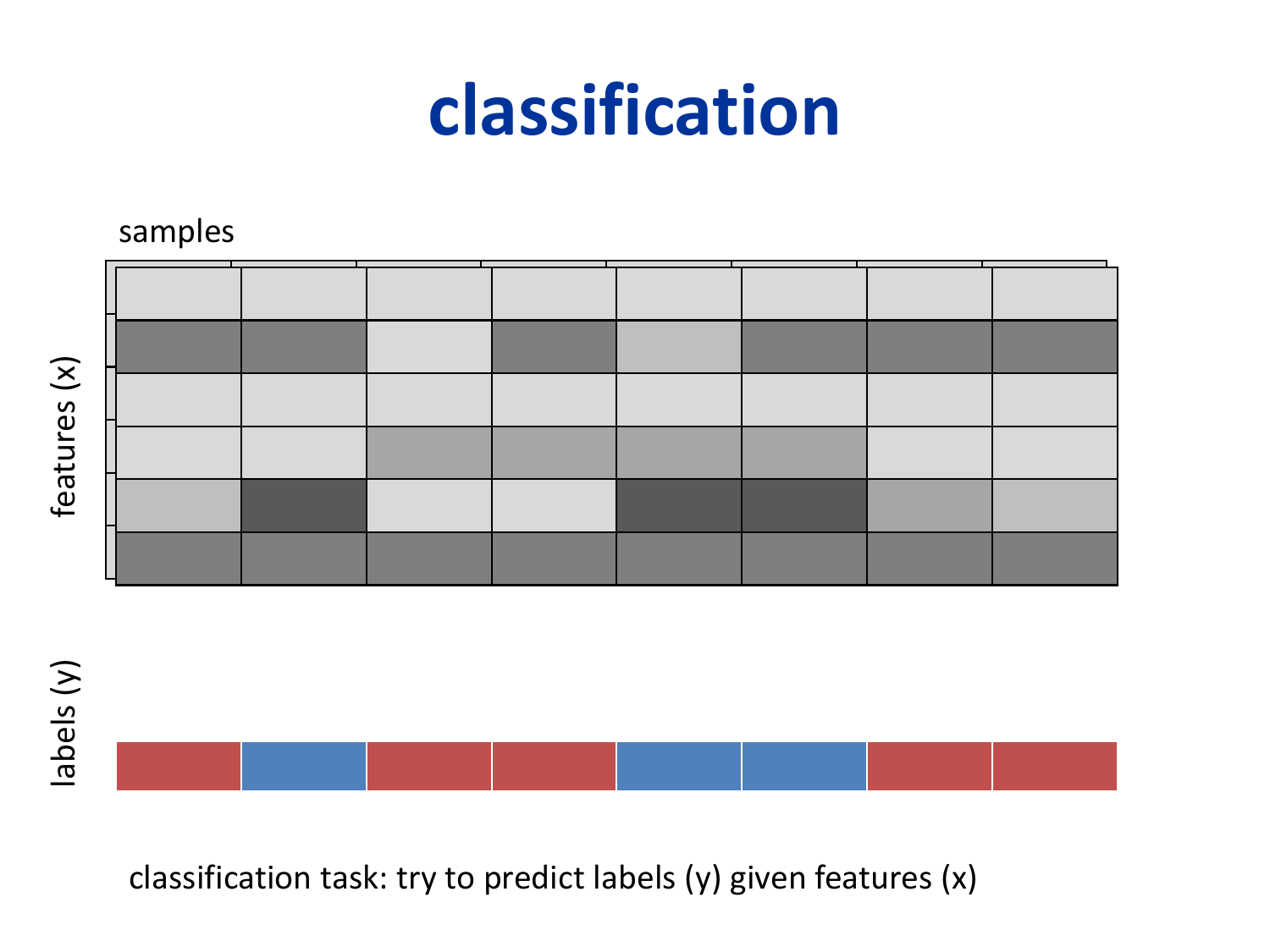}}
 \centerline{\includegraphics[width=0.9\linewidth]{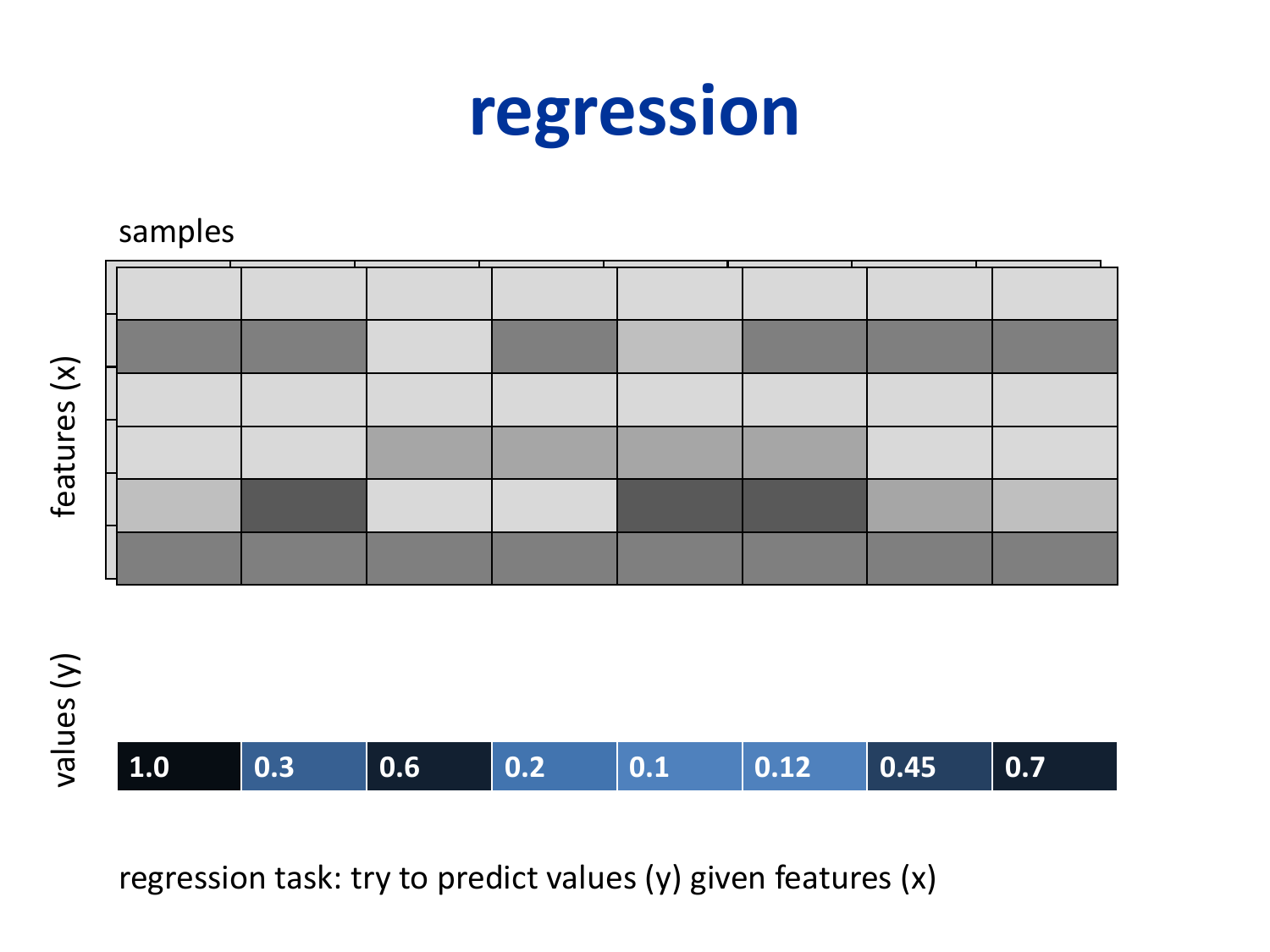}}
 \centerline{\includegraphics[width=0.9\linewidth]{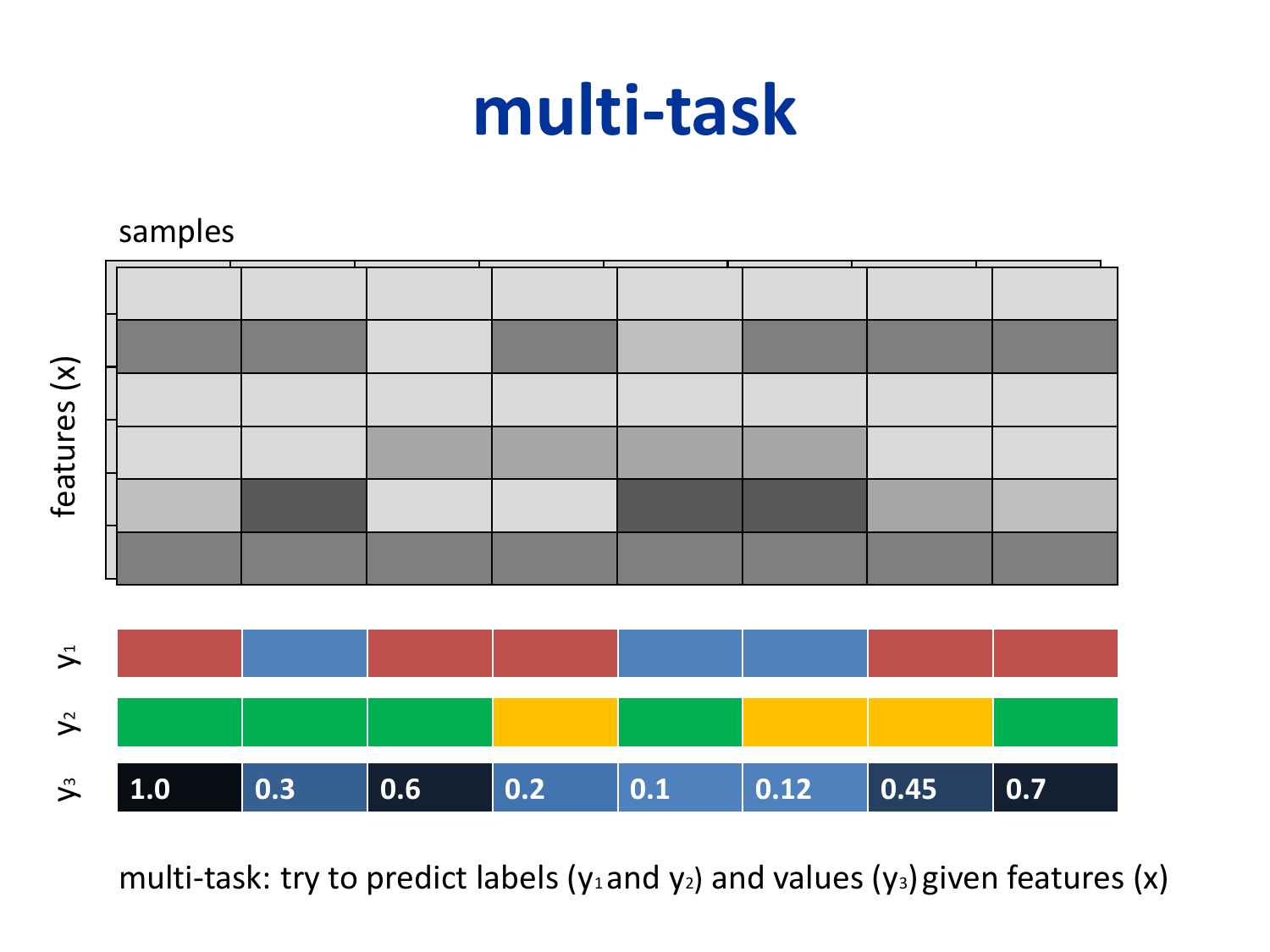}}
\noindent
\end{bgreading}

\begin{figure}[b]
\centerline{\includegraphics[width=\linewidth]{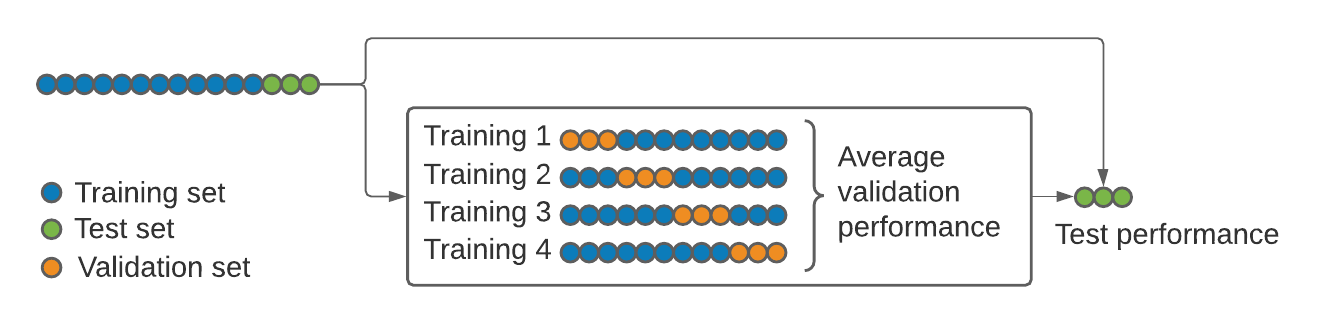}}
\caption{
N-fold cross-validation splits the dataset in N equally-sized subsets -- also known as stratification. For every training instance, one of the subsets is taken out of the dataset as a test set. The remaining samples constitute the training set. This set is used to train the model. The performance of the model trained on the training set is measured by comparing predictions of the samples in the test set to their actual values. This training is performed N times, until every subset has been used as a test set. Finally, the performances of all training instances are averaged to obtain a global measure of the model's performance. This also allows a standard deviation or confidence interval to be estimated.
}
\label{fig:ChSSPred:cross-validation}
\end{figure}

\subsection{Training and benchmarking structural property predictions}

Machine learning methods excel at finding patterns and relationships in data. However, for this they need to be trained on a big enough dataset of relevant samples: the \emph{training dataset}. This training process is done on labeled data, this means the predictor can `see' the correct class or value of the samples as it is already known. By this the predictor model can learn to recognize which features are associated with the prediction target. Thus, the machine learning model derives rules that capture how the input features relate to the output predictions, and those can then be used to make new predictions on (as-yet unlabeled) data.

An important aspect of training a supervised machine learning model is to estimate its performance while adapting during the training process. A dataset independent from the training data is needed for this, as reported performance in the training set may be inflated due to overfitting. A frequently used method in machine learning to estimate the performance is cross-validation, in which part of the training dataset is intentionally left out to be used as a \emph{validation dataset}, as shown in \figref{ChSSPred:cross-validation}. Furthermore, to measure the performance of the final machine learning model, a part of the available data should be completely kept out from the training process to be used afterwards as the \emph{test dataset}.

It is important to realise that close homologs will have high similarity in their sequences, as well as in their functions and structures. As a result, there is a danger that models become biased towards a certain sequence composition and are not representative of the full spectrum of sequence variation that may be encountered when applying the method to new data, and in particular to proteins that are not homologous to any proteins in the training data. Therefore, it is important to use a training and test set that do not contain (very) close homologs. In other words, the PDB structures should be filtered for sequence similarity before they can be used in model training \cite{Rost1993b}.

\begin{bgreading}[Learning non-local patterns]
\label{panel:ChSSPred:nonlocal}

Non-locality is an interesting characteristic for predicting structural properties from a sequence, especially due to long-range interactions. 

A sliding window can be used to collect values of a feature (e.g.\@ hydrophobicity) in a range of amino acids of a predetermined length to be used as input to the prediction model. 

Simplest prediction methods use some form of a sliding window, which only capture local patterns. Some older methods aimed to address this by using double (or nested) windows (see also \panelref{ChSSPred:history}). Most current methods use machine learning architectures such as convolutions or recurrent layers that can capture non locality directly.

Convolutions are the defining elements of Convolutional Neural Networks (CNNs). They consider a position's surroundings by processing its information through the application of kernels that extract specific patterns from the data. The values of a region of the input are multiplied by those in a kernel, summed and stored. Then, the filters are applied to an adjacent region of the input.
These steps are continued until all the positions of the input have been visited. It is the weights of the kernels that are learned in the training stage of such methods. The kernels can be thought of as units that can learn specific motifs. The creation of kernels is usually automated with most implementations, such as \code{pytorch}\footurl{pytorch.org/docs/stable/generated/torch.nn.Conv1d.html}.
Convolutions are often combined with other convolutions and pooling steps prior to passing their outputs into a neural network, which will predict a label.

Another way to predict long-range interactions is the recurrent units, the defining element of Recurrent Neural Networks (RNNs) \cite{Yu2019}. These units retain information from the context of a sequence in a trainable manner using a mechanism that is called \emph{gates}. In contrast with sliding window, information is not obtained up to a predefined sequence distance, but in a more flexible manner. Additionally, there are no pre-defined operations to be executed in each area, as is the case for CNNs. If a piece of information may be useful to improve a prediction, it may be obtained even from further away in the sequence. The gates may include some kind of `forget' function in most implementations, like the \code{pytorch} implementation of Long-Short Term Memory (LSTM)\footurl{pytorch.org/docs/stable/generated/torch.nn.LSTM.html}.

\end{bgreading}

\section{Sequence signatures}

Remember that the backbone parts of all naturally occurring amino acids are chemically identical (with the exception of proline, see \chref{ChIntroPS} \panelref{ChIntroPS:aas}). As secondary structure is stabilised by hydrogen bonding patterns between backbone atoms, the ability to form secondary structure is in essence a generic property of the peptide backbone. However, side chains have preferences for particular structural environments. Therefore, it is important to consider which patterns in the protein sequence are associated with specific structural elements. Those are the patterns that can serve as information sources from which structural property predictions can be made. Below we discuss some of these sources of information, and briefly put them in context of protein structures.

\subsection{Hydrophobicity patterns}
\begin{figure}[tb]
\centerline{\figlab{A}~\includegraphics[width=0.6\linewidth]{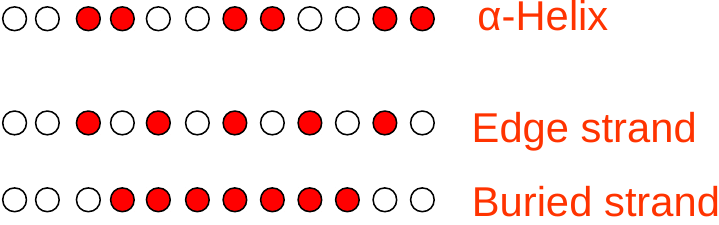}}
\centerline{\figlab{B}~\includegraphics[width=0.95\linewidth]{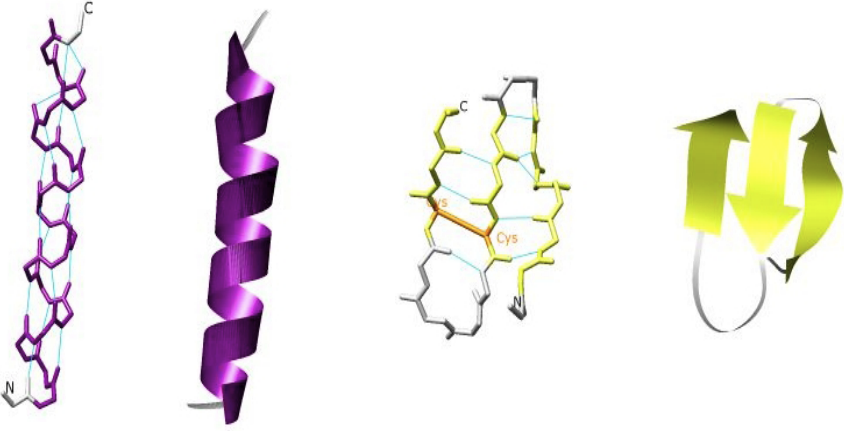}}
\caption{(A) Schematically and simplified, the hydrophobicity patterns in the sequence one may expect for different types of secondary structure elements; here, hydrophobic residues are indicated in red. (B) Examples of $\alpha$-Helical (left two) and $\beta$-strand (right two) structures. An $\alpha$-helix is often found at the protein surface, so that one side will be exposed to the solvent; this yields a sequence pattern of two hydrophobic, two hydrophilic residues, alternating. A $\beta$ strand will often be buried, with only the first and last residues hydrophylic; $\beta$ strands at the edge of the sheet, will have side chains alternately sticking `back' towards the protein (hydrophobic) and `out' into the solvent (hydrophilic).}
\label{fig:ChSSPred-ss-patterns}
\end{figure}

Sequence patterns of hydrophobicity can be very strong indicators for (secondary) structure types. For example, helices that are partially exposed to the solvent, will have hydrophobic residues on the buried side of the helix and polar or charged residues on the exposed side. This will lead to a periodic pattern of alternating hydrophobic and hydrophilic residues, with a period of (on average) 3.6 residues, see \figref{ChSSPred-ss-patterns}. For more background, please refer to \chref{ChIntroPS}. Similarly, a $\beta$-sheet with one side exposed to the solvent will show a sequence of alternating hydrophobic and hydrophilic residues, see \figref{ChSSPred-ss-patterns}. Loops are generally exposed to the solvent and therefore contain many more charged and polar residues than $\alpha$-helices and $\beta$-strands.

\subsection{Propensities}
\label{sec:ChSSPred:propensity}
Different structural environments lead to different likelihoods of observing certain amino acid types. This intrinsic preference of an amino acid for a certain structural environment is called \emph{propensity}. So, for example, you are less likely to find charged or bulky amino acids in a `buried' environment, and more likely to find hydrophobic amino acids on a protein-protein interaction interface. The propensity quantifies this difference in likelihood.

Generally, a propensity aims to reflect how much more likely a given amino acid is to be observed in a certain environment than randomly. Let's first introduce the fraction (or probability) of amino acids $p(total)_{S}$ in a particular structure type $S$, e.g.\@ the fraction of residues (in a protein) that are in an $\alpha$-helix: 
\begin{equation}
        p(total)_{S} = \frac{N(total)_{S}}{N(total)}, 
\end{equation}
where  $N(total)_{S}$ is the total number of residues in structure type $S$, and $N(total)$ is the total number of residues in the dataset. Let us furthermore consider the fraction of a specific amino acid type $aa$ in a particular structure type $S$: 
\begin{equation}
        p(aa)_{S} = \frac{N(aa)_{S}}{N(aa)},
\end{equation}
where $N(aa)_{S}$ is the number of amino acid type $aa$ in secondary structure type $S$, and $N(aa)$ is the total number of amino acid type $aa$ in all residue positions.

We can now calculate a propensity $P$ for amino acid type $aa$ for a specific type of structure $S$, by dividing the fraction of $aa$ found in $S$ ($p(aa)_{S}$) by the overall fraction of $S$ ($p(total)_{S}$), as follows:
\begin{equation}
        P(aa)_{S} = \frac{p(aa)_{S}}{p(total)_{S}}
\end{equation}

Note that the \emph{propensity} $P(aa)_{S}$ is not the same as the \emph{probability} $p(aa)_{S}$; propensity is a relative probability. When we calculate the propensity, we divide the fraction of residues of a specific amino acid in a secondary structure type by the total fraction of positions in that secondary structure type. Thus, a propensity below one indicates that an amino acid avoids that type of secondary structure, a propensity of around one indicates no preference, and a propensity larger than one indicates a (strong) preference of that amino acid for that secondary structure type. 

\subsubsection{Example: secondary structure propensities}
If 30\% of glutamic acid residues occur in an $\alpha$-helix, thus $p(Glu)_\alpha=0.3$, and 20\% of all residues are in an $\alpha$-helix, thus $p_\alpha=0.2$, then the propensity of glutamate for $\alpha$-helix becomes:\begin{equation*}
P(Glu)_\alpha = \frac{0.3}{0.2} = 1.5
\end{equation*}
So, in this example glutamate has a preference for the $\alpha$-helix.

\begin{figure}[b]
\centerline{
  \includegraphics[width=\linewidth]{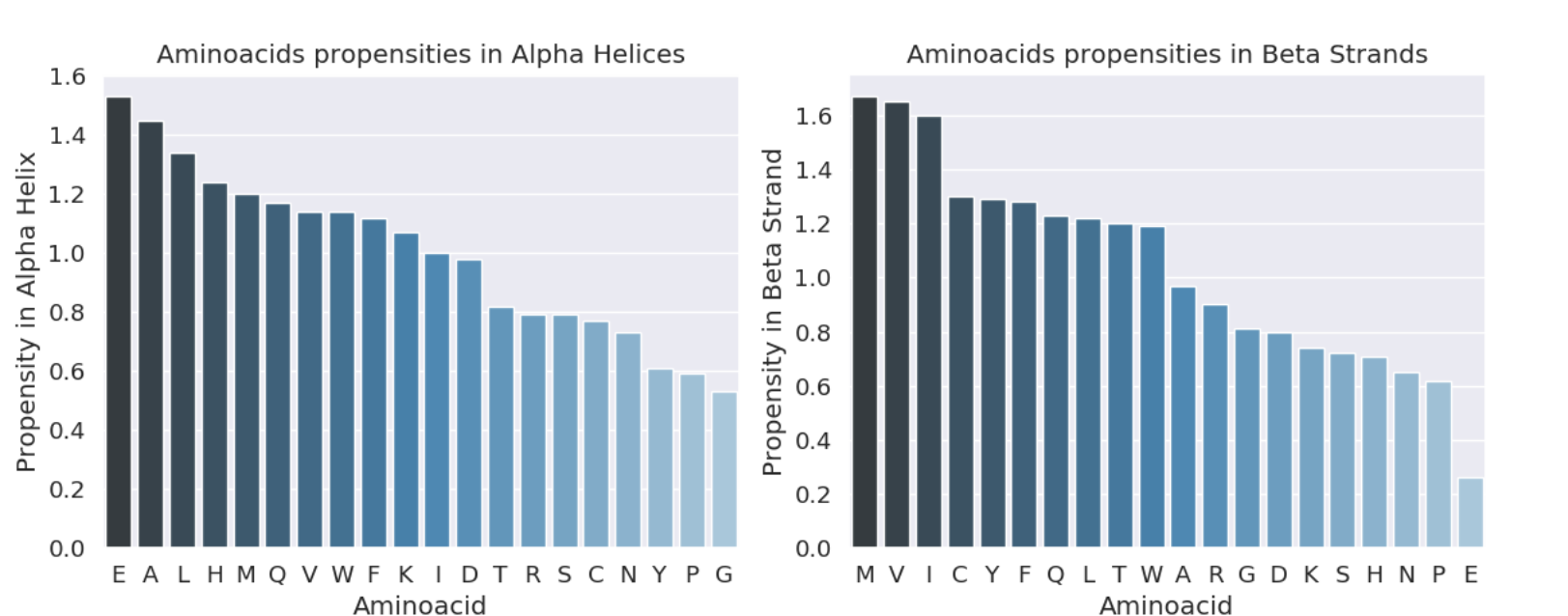}
}
\caption{Propensities of every amino acid type in $\alpha$-helix and $\beta$-strand. Based on data from \url{http://www.bmrb.wisc.edu/referenc/choufas.shtml}.}
\label{fig:ChSSPred-propensities}
\end{figure}

More generally, amino acids with side chains that are bulky close to the backbone -- more precisely, that have a branched structure at the C$\beta$ atom -- tend to favour $\beta$-strands, smaller amino acids tend to favour $\alpha$-helices and loops. Resulting propensities are shown in \figref{ChSSPred-propensities}.
Furthermore, residues with non-standard backbone configurations such as glycine and proline are often named `helix breakers' since they disrupt the helical pattern \cite{Aurora1998}, and may often be found at the ends (caps) of helices. Both residues, glycine and proline, are also enriched in loop regions, as they generally disrupt regular secondary structure patterns \cite{BrandenTooze,Imai2005}.

\begin{figure}[b]
\centerline{
  \includegraphics[width=0.7\linewidth]{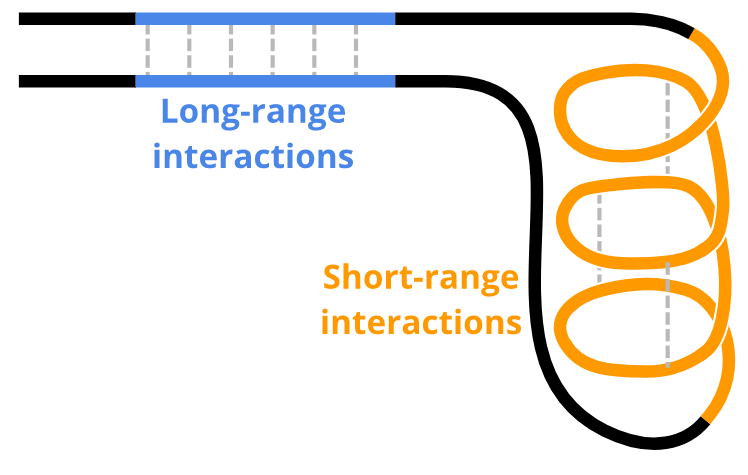}
}
\caption{Example of long-range and short-range interactions: within helical structure, interactions are always local (orange dotted lines, on the right); those between the strands in a sheet structure may be highly non-local (blue dotted lines, on the left).
}
\label{fig:ChSSPred-interaction-range}
\end{figure}

\subsection{Locality of (secondary) structure}
The interactions in an $\alpha$-helix are more local than those within a $\beta$-sheet, which have long-range interactions between residues on different strands as illustrated in \figref{ChSSPred-interaction-range}. Moreover, $\beta$-strands tend to be smaller continuous regions within the protein sequence compared to $\alpha$-helices, due to the extended conformation of the $\beta$-strand.  Overall, this makes $\alpha$-helices relatively easier to predict than $\beta$-sheets. In order to properly identify non-local interactions one has to take the protein-wide context into account. 

\begin{figure}
    \centerline{\includegraphics[width=0.7\linewidth]{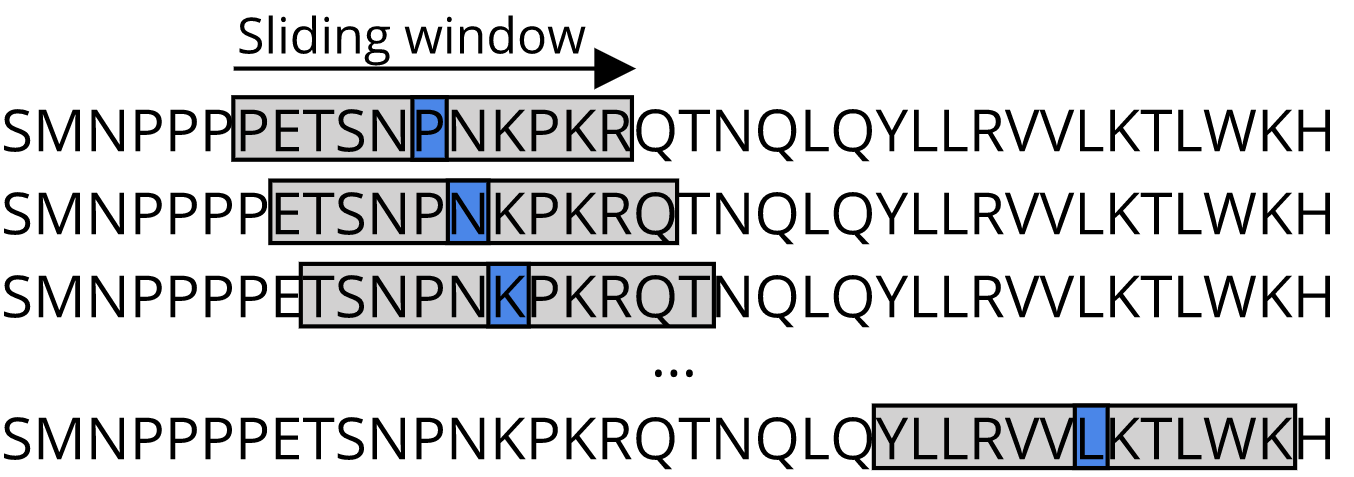}}
    \caption{The figure shows a window sliding along a protein sequence. For each amino acid it visits (blue), the values of the surrounding amino acids are used (grey).}
    \label{fig:ChSSPred-sliding-window}
\end{figure}

Early prediction methods suffered from a relatively poor performance when predicting $\beta$-strands because they only took the local sequence context into account. A sliding window approach, as shown in \figref{ChSSPred-sliding-window} can capture local patterns only. Incorporating non-local interactions in a prediction method, however, is far from trivial if we use window based approaches (see also \panelref{ChSSPred:history} below) \cite{Baldi1999,Magnan2014}. Recent progress in contact prediction (see \chref{ChHomMod} \secref{ChHomMod:contact-pred}) enables use of additional input for $\beta$-strand allocation, which can have a high impact on (secondary) structure prediction.

\subsection{Evolutionary information}
Sequence conservation patterns form a very strong indicator to recognise specific structural property types. All current state-of-the-art methods for structural property prediction take an evolutionary sequence profile as an input instead of a single sequence. Such profiles may be provided as the output of a multiple sequence alignment (MSA), a position-specific scoring matrix (PSSM) from BLAST or provided as a Hidden Markov Model (HMM) profile \cite{Woo2004,Jones1992}. All of these provide information on the probability of observing a residue type at a certain position. 
For example, in loop regions it is seven times more likely to have gaps (in the multiple alignment) than in an $\alpha$-helix or $\beta$-strand.

In the past several years, the use of representation models is causing much excitement. These models are based on the same architectures used for large language models. They are trained on huge amounts of unlabeled training data, and produce an internal representation which captures relevant evolutionary patterns. For any new input sequence, the model will generate a representation, typcally containing a thousand parameters which are then used as input features for a so-called downstream prediction model. This in practice often outperform methods based on MSAs, PSSMs or HMMs. While training these models is difficult, slow and expensive \citeeg{Capel2021}, once trained, their application is much faster than running a tool like BLAST or HMMer.

\section{Structural property prediction}
\subsection{Secondary structure}
Secondary structure (SS) prediction aims to label each residue in a protein to one of several secondary structure classes, typically $\alpha$-helix, $\beta$-sheet or coil. Note, this is distinct from the task of structure assignment, which is the task of assigning a structure label to each of the amino acids in a protein when the protein's three-dimensional (tertiary) structure is known, i.e.\@ we have the full atomic coordinates in the PDB. Secondary structure assignment can be viewed as a way to define a structure, and is thus used to create a benchmark or gold standard against which to evaluate the performance of SS prediction methods. In \chref{ChIntroPS} we discussed some programs that perform secondary structure assignment, such as DSSP \cite{Kabsch1983}, DEFINE \cite{Richards1988} and Stride \cite{Heinig2004}. In this section, we will focus on secondary structure prediction methods.
The first secondary structure prediction methods were developed during the 1970s, when only a handful of solved protein structures were available.

A very simple method for secondary structure prediction is to check which propensity is biggest for a given subset of sequence positions. We could do this is by looking at every residue in isolation, but this ignores the fact that the secondary structure of a residue is largely determined by its neighbours. A slightly more advanced approach is to average the propensity over a sliding sequence window \cite{Chou1978, Garnier1996}. We discuss several early methods that used this in the \panelref{ChSSPred:history}. Note that every residue is still only considered in isolation, so implausible configurations such as a single helical residue in isolation could still be predicted.

Increasing numbers of available protein structures in the PDB allowed for the training of neural net-based methods. Over time, severay types of architectures have been successful, such as (artificial) neural networks (ANN) \cite{Rost1993b}, recursive neural nets (RNN) \cite{Pollastri2002, Baldi2003} and convolutional neural networks (CNN) \cite{Wang2016a}. An 
RNNs are able to capture similar trends as HMMs. 
Which model can be used best depends on the dataset, the prediction task and the aim of the study. Besides, different methods may be combined into one architecture.

The most commonly observed secondary structures were introduced in \chref{ChIntroPS}. Classical secondary structure prediction methods have typically approached it as a three-state classification task: the problem of affixing one of three labels, $\alpha$-helix, $\beta$-sheet or coil, to each residue in a protein sequence.

Traditionally, secondary structure prediction has been approached as a classification task. More recent methods such as NetSurfP2.0 \cite{Klausen2019} have started to include prediction of other attributes, such as surface accessibility and backbone dihedral angles, which are regression tasks. Other methods such as PSIPRED \cite{Jones1999, Buchan2019} treat secondary structure as a regression task, and predict propensities of an amino acid to be in sheet, coil or helix conformation. 

\begin{bgreading}[History of secondary structure prediction]
\label{panel:ChSSPred:history}
Many different machine learning algorithms have been developed to tackle the problem of structural property prediction. Many of these methods incorporate evolutionary relationships by creating sequence profiles using \mbox{(PSI-)BLAST}, which are then fed into the model as training data \cite{Rost1993b,Woo2004}. Besides, evolutionary information can be stored in Hidden Markov Model (HMM) profiles, which also includes position dependent penalties for amino acid deletions and insertions \cite{Eddy1996, Bystroff2008}. 

Early methods window-based approaches are the Chou-Fasman algorithm \cite{Chou1978} and 
the GOR family of algorithms \cite{Garnier1996}, see for more details the review by \citet{Pirovano2010}. Other simple prediction models were used to make structural property prediction, e.g.\@ multiple sequence alignment~\cite{Frishman1996}, k-nearest neighbor~\cite{Salamov1995}, decision trees~\cite{Rost1994}, and support vector machines~\cite{Hua2001}. For many structural prediction tasks, like secondary structure prediction, it is beneficial to include non-linearity. Different approaches have been developed, e.g.\@ by identifying correlations that indicate $\beta$-strand connectivity. These correlations can be used to strengthen the signal to predict the existence of a $\beta$-sheet in the sequence. For example, Predator uses weak sequence signals that indicate correlations between (contacting) residues in adjacent $\beta$-strands \cite{Frishman1996}. Propensity values for hydrogen bonding in a sliding window are used  to predict $\beta$-strands. Other examples include PHD, which uses homologous sequences identified by BLAST to incorporate evolutionary information and SSPro, which uses three sliding windows to identify possible $\beta$-strand interactions \cite{Rost1994, Pollastri2002}. 

One of the earliest sophisticated methods - YASPIN~\citep{lin2005simple} utilised evolutionary profiles, HMMs and a neural network architecture with a slightly different strategy. The advantage of the YASPIN method was high speed mostly due to its simplicity: Instead of using an alignment algorithm directly, YASPIN method applied a 15-residue PSSM window generated from PSI-BLAST. The 7-state (instead of the common 3-state) secondary structural output was generated by the neural network in order to obtain more information that was afterwards filtered by an HMM to ultimately output a 3-state secondary structures. The major strength of the method was its ability to predict b-strands with high accuracy. 
 
\end{bgreading}

Recent methods use multi-task prediction architectures which are trained on multiple tasks simultaneously. 
The aim is improve a specific structural prediction task by adding related prediction tasks \cite{Caruana1997}. 
In previous methods, structural annotations for specific tasks are often used as input features for the prediction of the main task. Learning all tasks at the same time can transfer the same information and requires fewer pre-processing steps. 
Such architectures predict secondary structure in combination with other structural properties such as solvent accessibility, disorder, backbone angle and residue contacts \cite{Pollastri2002,Heffernan2015,Capel2022Multi-TaskPrediction}. 
The two leading examples are NetSurfP2.0 \cite{Klausen2019} and OPUS-TASS \cite{Xu2020}.

Another recent developed model that aims to to transfer information between multiple tasks and has received a lot of interest is the Transformer model \cite{Devlin2019,Rao2019,Vig2020}. Importantly, these deep learning models require a substantial amount of training data that is becoming more easily approachable by the increasing amount of protein structural data \cite{Capel2021}.

\subsection{Coiled coil}
Coiled-coil is a pair of helices that together form a twisted rod and are typically DNA binding (\chref{ChIntroPS} \figref[e]{ChIntroPS:atypical-ss}). Due to the twist in the coiled-coil, there is a seven-residue repetitive element where residues of both helices are in direct contact. Typically, a leucine is found at each seventh residue, and in between (at each third and fourth residue), a valine or isoleucine. Due to the repeating leucine, these structures are also known as `leucine zippers' or `leucine-rich repeats'. This pattern makes them fairly easy to detect and predict \cite{Lupas1997}. COILS \cite{Lupas1997}, a profile based methods was the first coiled coil prediction algorithm developed. Later, HMM based methods like Marcoil \cite{Delorenzi2002} were developed which improved predictions for short coiled coil regions. Recent methods like DeepCoil \cite{Ludwiczak2019} use deep learning and have higher sensitivity and accuracy than profile based and HMM based methods.

\subsection{Surface accessibility}
Among various structural properties, surface accessibility predictions of amino acids is of major importance. Residues that are exposed to the environment can have many different functions. For example, they can be part of the catalytic site of an enzyme or participate in PPIs. Furthermore, knowing which residues are on the surface of a protein can be important in drug design, e.g.\@ in molecular docking \cite{Ferreira2015,Naderi-manesh2001}.

There are two ways to approach the problem of surface accessibility prediction: as classification or as regression (\panelref{ChSSPred:ML}). Classification methods predict whether a residue is buried, exposed or partially exposed based on a threshold and do not regard the absolute value of surface exposure \cite[e.g.][]{Naderi-manesh2001,Ahmad2003}. Regression methods, on the other hand, aim to predict which fraction of a residue is exposed \cite[e.g.][]{Petersen2009, Wagner2005}.

Because secondary structure prediction and solvent accessibility prediction are methodologically similar problems, several methods have been developed that aim to tackle both \cite{Heffernan2015, Klausen2019}. Finally, solvent accessibility prediction can also be used to improve the accuracy of other structural property predictions and vice versa \cite{Faraggi2012a, Klausen2019}.

\label{sec:atypical}

\subsection{Disorder and flexibility} 
Disordered proteins or protein regions (\figref[c,d]{ChIntroPS:atypical-ss}), are those that lack a folded structure. Disorder prediction is relatively easy compared to 3D structure prediction. A simple but effective approach is to count amino acids with high propensities for disorder \cite{Oates2012} -- these are charged and polar (hydrophilic) amino acids -- inside a sliding window over the sequence. There are more advanced predictors which use hidden Markov models (HMMs) \cite{Cheng2005}  and Recurrent Neural Networks (RNN), for example, DisoMine \cite{Orlando2018}. Recently, the Critical Assessment of protein Intrinsic Disorder prediction (CAID) experiment was designed to assess the prediction methods for intrinsic disorderd proteins (IDPs) \cite{Necci2021}. Deep learning methods like RawMSA outperformed the physicochemical based methods in the first CAID experiment \cite{Mirabello2019}.

Flexibility is related to disorder, but not necessarily the same \cite[e.g.][]{Pancsa2016}. Protein flexibility influences a protein's biological activity like catalysis and stability. DynaMine is a dedicated method that aims to predict backbone and sidechain flexibility from sequence \cite{Cilia2014,Cilia2013, Raimondi2017}.%

\subsection{Transmembrane regions} 
Transmembrane (TM) proteins exist inside a membrane environment which is largely non-polar. Therefore, the membrane-spanning region of the TM protein will tend to have amino acids with hydrophobic side chains \emph{on the outside} to match the apolar lipid environment of the membrane. In case of pore or channel proteins, the inside (enclosed by a `ring' of helices or a $\beta$-barrel) will tend to be hydrophilic \cite{Krogh2001}. Early methods in transmembrane topology prediction were based on hydrophobicity analysis \cite{Yuan2004}. Other methods for the prediction of TM helices utilize machine learning approaches including HMM \cite{Krogh2001,Tusnady1998}) or SVM \cite{Yuan2004}. For the prediction of TM $\beta$-barrels, neural network approaches have been developed \cite{Jacoboni2001,Gromiha2007}. 

\subsection{Aggregation propensity} 
Some proteins can aggregate into specific insoluble $\beta$-stranded structures called amyloid fibrils (\chref{ChIntroPS} \figref[a,b]{ChIntroPS:atypical-ss}). Early results show that several proteins associated with disease also have a high propensity for amyloid fibril formation \cite{Chiti2006}. There are multiple amyloid fibril prediction tools \cite{Zibaee2007}, however, reference databases are still small, making it difficult to validate such methods \cite{Micsonai2015}. Since protein aggregation is mostly linked to amyloid fibrils with cross-$\beta$ structure, various algorithms have been developed to predict aggregation-prone parts from the primary sequences. PASTA, for example, is a protein aggregation predictor that was trained on a dataset of globular proteins of known native structure and predicts propensities of two residues to be a part of a cross-beta structure of neighbouring stands \cite{Walsh2014}.

\section{Practical advice}
In the previous sections multiple methods for structural property prediction are discussed. In this section, we will provide some tips for end users of structural property prediction algorithms: 
\begin{compactitem}
\item Firstly, check the recent literature for the latest best-performing methods, preferably from a review where all methods have been vetted in the same way on the same dataset.
\item If possible, benchmark some of the best-scoring tools on relevant cases to get an idea of their accuracy for your purposes.
\item Finally, check the similarity between the different methods, but also the similarity in prediction of specific regions. Methods or regions that get the same prediction with very different methods are generally the most reliable ones. 
\end{compactitem}

\begin{bgreading}[Caveats]
\label{panel:ChSSPred:caveats}

Using the most accurate structural property prediction methods, predict up to 80\% of residues correctly. Is it possible to do better still or is this close to the maximum attainable performance? There are a number of fundamental reasons why it is difficult, if not impossible, to further improve predictive performance:

\begin{itemize}
\item[\textbf{Biases in the reference set will affect predictions.}] A large fraction of the structures was experimentally solved using X-ray crystallography. This process, which requires a stable protein conformation to succeed, may lead to biases towards more stable structures. For example, a region that is disordered under normal conditions may be removed completely or be stabilized as e.g\@ a $\beta$-sheet in crystal form. Theoretically, NMR structure determination should suffer less from these problems as the proteins are measured in solution, but the heuristic algorithms used to find the most plausible conformations may still lead to biases.

\item[\textbf{The native state is not static}]
A protein in solution is not fixed in its native conformation, but shows significant internal motion. For example, globular proteins continuously switch between the native and unfolded conformations. Many transmembrane proteins have big conformational changes essential to their function (e.g.\@ they may change conformation upon binding of an ion that allows the ion to pass through the membrane). Furthermore, the (secondary) structure may not even be stable in the native state. For example, a region may show constant transitions between a disordered state and a metastable helical state \cite{Linding2003,Kagami2021,Kagami2021a}.

\item[\textbf{A protein molecule does not exist in isolation.}] Many
proteins are post-translationally modified in a way that may induce a change in conformation \cite{Xin2012}. Other examples of environmental interactions include binding to other molecules after which a particular conformation is stabilized or even a conformational transition based on the acidity or temperature of the environment. When we try to predict the structural properties from sequence we take none of such factors into account.

\end{itemize}

\end{bgreading}

\section{Key Points}

\begin{compactitem}
\item Structural property prediction can be solved well with machine learning methods.
\item High accuracy methods ($80\%-90\%$) are available for secondary structure, solvent accessibility, disorder, transmembrane regions and aggregation propensity.
\item Recent and best performing methods use representation models as input feature, and are typically trained on  multiple properties (multi-task)
\end{compactitem}

\section{Further Reading}

\begin{compactitem}
\item Secondary structure patterns -- \citet{BrandenTooze}, in particular Chapter 2 ``Motifs of Protein Structure''
\item ``Biological Sequence Analysis'' -- \citet{Durbin}
\item Review on secondary structure prediction methods -- \citet{Pirovano2010}
\item ``Ten quick tips for sequence-based prediction of protein properties using machine learning'' -- \citet{Hou2022}
\end{compactitem}

\section*{Author contributions}
{\renewcommand{\arraystretch}{1}
\begin{tabular}{@{}ll}
\ACtxt: &   MD, PB, KW, DG, JG, JvG, SA, JH, KAF \\
\ACfig: &   JG, JvG, JH \\
\ACref: &   JH, MD, JG, MO \\
\ACproof:&  BS, JvG, SA, KAF \\
\ACfb:  &   MO, RB, IH \\
\ACeds: &   JvG, SA, KAF
\end{tabular}}

\noindent
The authors thank \TL~\TLid{} and \CG~\CGid{} for critical proofreading.

\mychapbib
\clearpage

\cleardoublepage

\end{document}